\documentclass{ppai}

\usepackage{ppai}

\usepackage[utf8]{inputenc} 
\usepackage[T1]{fontenc}    
\usepackage{hyperref}       
\usepackage{url}            
\usepackage{booktabs}       
\usepackage{amsfonts}       
\usepackage{nicefrac}       
\usepackage{microtype}      
\usepackage{lipsum}
\usepackage{graphicx}
\graphicspath{ {./images/} }
\usepackage{xcolor}
\usepackage{algorithm}
\usepackage{algpseudocode}
\usepackage{amsmath}

\usepackage{enumitem}
\usepackage{soul}

\algrenewcommand\algorithmicrequire{\textbf{Input:}}
\algrenewcommand\algorithmicensure{\textbf{Output:}}

\newcommand{\pCOMB}{\ensuremath{\pi_{\mathsf{CONCAT}}}}
\newcommand{\pPRE}{\ensuremath{\pi_{\mathsf{PREPROCESS}}}}
\newcommand{\pREVPRE}{\ensuremath{\pi_{\mathsf{REV-PREPROCESS}}}}
\newcommand{\pEVAL}{\ensuremath{\pi_{\mathsf{EVAL}}}}
\newcommand{\pSDG}{\ensuremath{\pi_{\mathsf{SDG}}}}
\newcommand{\pAVG}{\ensuremath{\pi_{\mathsf{AVG}}}}
\newcommand{\pBIN}{\ensuremath{\pi_{\mathsf{BIN}}}}
\newcommand{\pLR}{\ensuremath{\pi_{\mathsf{LR}}}}
\newcommand{\pWLE}{\ensuremath{\pi_{\mathsf{WLE}}}}
\newcommand{\pEQ}{\ensuremath{\pi_{\mathsf{EQ}}}}
\newcommand{\pMUL}{\ensuremath{\pi_{\mathsf{MUL}}}}
\newcommand{\pNM}{\ensuremath{\pi_{\mathsf{NOISY-MARG}}}}

\newcommand{\pSORT}{\ensuremath{\pi_{\mathsf{SORT}}}}
\newcommand{\pLT}{\ensuremath{\pi_{\mathsf{LT}}}}
\newcommand{\pDIV}{\ensuremath{\pi_{\mathsf{DIV}}}}
\newcommand{\pINVBIN}{\ensuremath{\pi_{\mathsf{INV-BIN}}}}
\newcommand{\pGUASS}{\ensuremath{\pi_{\mathsf{GAUSS}}}}
\newcommand{\pRDM}{\ensuremath{\pi_{\mathsf{GR-RANDOM}}}}
\newcommand{\pSOFTMAX}{\ensuremath{\pi_{\mathsf{SOFTMAX}}}}
\newcommand{\pABS}{\ensuremath{\pi_{\mathsf{ABS}}}}

\title{End to End Collaborative Synthetic Data Generation}

\author{
  Sikha Pentyala, Geetha Sitaraman, Trae Claar, Martine De Cock \\ 
  University of Washington Tacoma\\ 
  \texttt{\{sikha,sgeetha,tclaar,mdecock\}@uw.edu} 
}

\begin{document}
\maketitle
\begin{abstract}
The success of AI is based on the availability of data to train models. While in some cases a single data custodian may have sufficient data to enable AI, often  multiple custodians need to collaborate to reach a cumulative size required for meaningful AI research. The latter is, for example, often the case for rare diseases, with each clinical site having data for only a small number of patients. Recent algorithms for federated synthetic data generation are an important step towards collaborative, privacy-preserving data sharing. Existing techniques, however, focus exclusively on synthesizer training, assuming that the training data is already preprocessed and that the desired synthetic data can be delivered in one shot, without any hyperparameter tuning. In this paper, we propose an end-to-end collaborative framework for publishing of synthetic data that accounts for privacy-preserving preprocessing as well as evaluation. We instantiate this framework with Secure Multiparty Computation (MPC) protocols and evaluate it in a use case for privacy-preserving publishing of synthetic genomic data for leukemia.    
\end{abstract}


%
%

\section{Introduction}\label{sec:intro}
The transformative potential of AI in important domains such as healthcare and finance is often hindered by limited access to high-quality, realistic data. Although vast amounts of data exist, they remain locked up in silos, controlled by entities such as hospitals or banks, guarded by privacy regulations.
Researchers must go through lengthy and cumbersome administrative processes to access sensitive data, limiting their ability to conduct impactful research \cite{watson2023delivering}. 

Synthetic data generation (SDG), i.e.~the generation of artificial data using a 
synthesizer trained on real data, offers an appealing solution to make data available while mitigating privacy concerns \cite{hu2023sok}. When done well, synthetic data has the same characteristics as the original data but, crucially, without replicating personal information. This makes it suitable for open publication and fostering AI research. An important characteristic of synthetic data is that the generated data fits any data processing workflow designed for the original data. This means that synthetic data can be used by anyone with basic data science skills in the comfort of their preferred data analysis software, even if they have no technical knowledge about privacy-enhancing technologies (PETs).
As a result, synthetic data has become a common technique to fuel data science competitions, and SDG is increasingly establishing itself as the way forward to broad privacy-preserving data sharing in practice.


Data custodians with sufficient data can often train their own synthetic data generators and publish synthetic datasets. In many real-world scenarios, however, individual data custodians often lack enough representative data to generate high-quality synthetic datasets on their own \cite{prediger2024collaborative}. The 2023 U.S-U.K.~PETs Prize Challenge, for instance, used synthetic data for financial fraud detection that was generated based on 
real data from the global financial institutions BNY Mellon, Deutsche Bank, and SWIFT.\footnote{\url{https://petsprizechallenges.com/}} Another common use case are rare diseases, with each data custodian (clinical site) having data for only a few patients \cite{Claar:2024}.

In the case of the above mentioned PETs Prize Challenge, the original real data was brought together and disclosed to the company Mostly AI for synthetic data generation. Such reliance on a trusted third party is not always desirable or even legally possible. Broad adoption of privacy-preserving data sharing across data silos requires a collaborative approach to SDG with both input and output privacy guarantees. Output privacy, commonly achieved through Differential Privacy (DP) \cite{dwork2006calibrating}, means that the synthetic data does not leak sensitive information about the real data that was used to train the synthesizer. Input privacy, which is particularly relevant in the collaborative setting, ensures that none of the data custodians need to disclose their raw data to any other entity.

\vspace{0.5em}
\noindent
\textbf{Problem.} Existing research has explored methods for collaboratively generating privacy-preserving synthetic data from multiple data custodians while ensuring both input and output privacy. Approaches based on federated learning (FL) \cite{maddock2023flaim} rely on a central server or a trusted entity, and yield synthetic data with lower utility than in the centralized setting with global differential privacy (DP).\footnote{Global DP refers to applying DP to achieve output privacy in a centralized setting, where all data from multiple custodians is available in one location with a trusted entity.} To address this limitation, Pentyala et al.~\cite{pentyala2024caps} proposed leveraging cryptographic techniques such as Secure Multiparty Computation (MPC) \cite{CDN2015} to generate synthetic data, mimicking global DP and producing high-quality data, at the cost of increased runtime. \textit{What is common across all these existing solutions is that they focus exclusively on synthesizer training. They assume that the training data has already been preprocessed, and they are limited to one-shot synthesizer training and data publication.}
Generating high-quality synthetic data, however, often involves a multi-stage process, including data preparation, evaluation of the synthetic data against real data, and hyperparameter tuning for the SDG algorithm.

\begin{itemize}[leftmargin=*,topsep=1.5pt,itemsep=1.5pt]
\item \textit{Preprocessing.} 
Many SDG algorithms require specific preprocessing steps to generate synthetic data. For instance, SDGs that follow the select-measure-generate paradigm, which are shown to be effective for tabular datasets, often rely on categorical data, hence requiring discretization of continuous features \cite{tao2021benchmarking}. In existing work on federated SDG this is handled by performing discretization on the dataset as a whole before distributing it across silos \cite{maddock2023flaim}, or by using a simple discretization algorithm that does not require knowledge about the data \cite{pentyala2024caps}. The former is unrealistic in practical scenarios where the data originates from different sources and cannot be brought together -- the very scenario that federated SDG is supposed to address! In case of binning, for example, the latter is done with equiwidth binning, in which one assumes that the range (potential minimum and maximum values) of the continuous feature are known, and one subsequently divides this range into a fixed number of intervals that are each of the same length. Depending on the data distribution, this can lead to inferior results compared to equidepth binning, such as quantile binning, in which interval boundaries are chosen dynamically based on the data such that each interval (bin) contains approximately the same number of instances. 

Preprocessing techniques in general can yield better results 
for downstream tasks, such as training of synthetic data generators,
when performed on combined data from multiple data custodians, rather than having each data custodian preprocess their datasets locally, as is typically done in FL. One potential approach is to perform privacy-preserving preprocessing with DP guarantees, such as quantile binning or normalization, over the combined data. An advantage of this is that one can release DP statistics, such as DP quantiles or means, which are used to convert generated synthetic data back into a format that resembles real data. This process of DP preprocessing consumes part of the privacy budget.

\item
\textit{Evaluation}
Evaluating synthetic datasets against real data is a crucial step in assessing the quality of the generated synthetic data. The evaluation results, i.e., evaluation metrics, help in refining the SDG process by guiding the hyperparameter tuning or retraining of the SDG model. Furthermore, evaluation metrics are often used to decide whether the generated synthetic data meets the quality standards necessary for publication. Many a times evaluation metrics are published before publishing synthetic data, which has been shown vulnerable to privacy attacks \cite{ganev2023inadequacy}. 
Some FL frameworks allow local evaluation of synthetic data against local real data during each FL round but require publishing evaluation metrics or some DP information, which consume part of the privacy budget.

\item
\textit{Hyperparameter tuning.}
Many proposed privacy-preserving frameworks for generating synthetic data over combined datasets do not focus on hyperparameter tuning of the SDG model. Their primary focus is on publishing the SDG model or synthetic data at once. However, hyperparameter tuning often requires multiple iterations of evaluation, synthetic data generation, and potentially preprocessing (if the dataset changes with each iteration). Each run of the entire pipeline consumes privacy budget, specifically when either synthetic data, generator, or the evaluations metrics or any other information is published. This can severely impact the quality of the final published synthetic data. To address this, it is essential to publish only high-quality synthetic data at the end, while minimizing the privacy budget spent. 
\end{itemize}

The above considerations call for a framework capable of \textit{performing privacy-preserving preprocessing across data silos, and publishing the synthetic data only after parameter tuning and ensuring that desired quality standards are met.} To the best of our knowledge, there is no work in the open literature that performs \textit{the entire pipeline of SDG on data from multiple data custodians while preserving input privacy and conserving the privacy budget}.

\vspace{0.5em}
\noindent
\textbf{Related Work.}
While the primary focus of existing research has been on generating  synthetic data while providing input and (or) output privacy guarantees \cite{pentyala2024caps, shankar2024silofuse,zhao2023libertas,xin2022federated, lomurno2022sgde}, little focus has been on achieving hyper parameter tuning for optimal SDG process. Current literature focuses on efficient DP parameter tuning in a centralized setting \cite{koskela2024practical, xiang2024does, ding2022revisiting}.  Mitic et al. propose PrivTuna for effective hyperparamter tuning in cross-silo federated setting\cite{mitic2024}. PrivTuna leverages multiparty homomorphic encryption to share the locally tuned parameters and performance metrics. There is no literature, to the best of our knowledge, that performs privacy-preserving hyperparameter tuning and runs the entire SDG pipeline while providing both input and output privacy.

\begin{figure}[ht!]
    \centering
    \includegraphics[width=\linewidth]{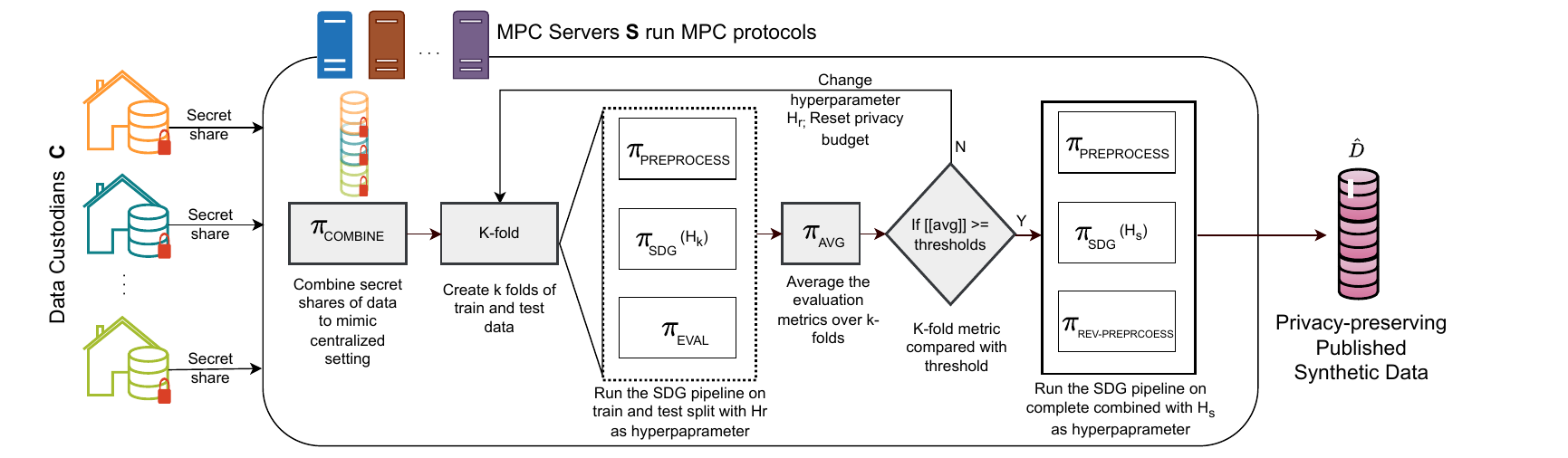}
    \caption{Proposed framework for privacy-preserving publishing of synthetic data}
    \label{fig:framework}
\end{figure}

\vspace{0.5em}
\noindent
\textbf{Our Contribution.} Our goal is to generate synthetic data of a desired quality level, based on real data from multiple data custodians while preserving input privacy. We aim to achieve this through privacy-preserving hyperparameter tuning and by publishing only the final generated synthetic dataset, while consuming as little privacy budget as possible. The core idea of our proposed method is to iteratively run the entire SDG pipeline while evaluating the quality of the generated data at each iteration while disclosing as little information as possible throughout. Each iteration starts with a set of hyperparameter values and allots privacy budget, say $\epsilon$, for the preprocessing and training of the SDG model. For simplicity, we propose to leverage k-fold evaluation to select the optimal set of hyperparameters in each iteration \footnote{This can be extended to sophisticated hyperparamter selection algorithms such as Optuna \cite{akiba2019optuna}. Note that the focus of the paper is introducing a framework for the entire pipeiline with input privacy.}. After evaluating the synthetic data, if the averaged evaluation metrics do not meet the specified thresholds, we loop over the process. We reset the privacy budget to $\epsilon$ for each iteration since \textit{we are not revealing or publishing any of the outputs of the SDG pipeline yet} - synthetic data, SDG model, or metrics. This helps avoid unnecessary consumption of the privacy budget. Once a suitable set of hyperparameters is identified for the SDG model, we preprocess and retrain the model on the entire dataset within the same privacy budget, $\epsilon$, and only then we publish the dataset or the generator. In this approach, although multiple iterations of the pipeline are run, the total consumed privacy budget remains $\epsilon$, the same as spent in just one run of the SDG pipeline. The above is easy to adapt in a centralized setting, but is challenging to adapt to cross-silo settings.

To extend the above method for multiple data custodians while preserving input privacy, we build upon the method proposed by Pentyala et al.~\cite{pentyala2024caps}, which replaces the trusted central entity of the centralized paradigm with a set of MPC servers. These MPC servers perform computations over secret-shared data, ensuring that the computing servers never see the private data of the custodians, thus preserving input privacy. Their method applies DP in the MPC context (DP-in-MPC), i.e.,~generation\footnote{MPC servers generate secret shares of noise.} and addition of noise that satisfies DP is done within MPC. This approach has the advantage of achieving utility results similar to a centralized setting, and it works regardless of how the data is partitioned across the data silos. Specifically, we propose running the entire SDG pipeline as described above using MPC (see Algorithm \ref{alg:overview}). Our framework is designed to be modular -- it requires DP-in-MPC protocols for preprocessing, MPC protocols for evaluation, and DP-in-MPC protocols for training the SDG model as shown in Figure \ref{fig:framework}. An important advantage of our framework is its asynchronous nature, allowing data custodians to secret-share their data and delegate the process to the MPC servers. 

We demonstrate the effectiveness of our proposed framework by generating synthetic genomic data for leukemia\footnote{Leukemia is a type of blood cancer that originates in the bone marrow. We generate synthetic data for different disease types, namely ALL, AML, CLL, CML, and ``Other''.} based on data from multiple hospitals. Our work and research on designing optimized MPC protocols and generating synthetic genomic datasets for rare diseases using the mentioned framework, is in progress. As a preliminary demonstration, we analyze the framework for runtime with naive MPC protocols. We, additionally, assess the quality of the generated synthetic data  compared to the real combined data with multiple evaluation metrics.
Note that this is not part of the framework and will not be used in practice; it is solely for the purpose of empirical evaluation. We summarize our main contributions as below:
\begin{itemize}[leftmargin=*,noitemsep,topsep=0pt]
    \item We are the first to propose a framework to publish DP synthetic data, based on the data from multiple silos, that runs the entire pipeline of SDG, including hyperparameter tuning, while providing input and output privacy, and minimizing the consumption of privacy budget.
    \item The proposed framework leverages DP-in-MPC to yield synthetic data with a level of fidelity and utility  similar to the centralized paradigm. Furthermore, it works for any kind of data partitioning.
    \item We demonstrate the framework by generating synthetic genomic data for leukemia. 
    \item For the above, we build (naive) MPC protocols (a) to preprocess leukemia data across data silos using quantile binning (b) to evaluate the generated data in terms of utility (with logistic regression) and in terms of fidelity (workload error) with k-fold cross-validation (c) MPC protocols for synthetic data generation.
\end{itemize}

\section{Preliminaries}\label{sec:prelims}
\textbf{Secure Multi-Party Computation (MPC).} MPC refers to a class of cryptographic approaches that enable multiple distrustful parties to privately compute a function without revealing their private inputs to any other entity \cite{cramer2000general,evans2018pragmatic}. We consider secret sharing based MPC protocols. Our framework employs the ``MPC as a service'' paradigm, where the data custodians delegate their private computations to a set of non-colluding servers, which we refer to as MPC servers. Each data custodian splits their private input $v$ into secret shares denoted as $[\![v]\!]$, and sends the shares to the MPC servers, following the given MPC scheme. Though $v$ can be reconstructed when the shares are combined, none of the MPC servers can individually learn the value of $v$. The MPC servers then proceed to jointly perform computations over the secret-shared data such as preprocessing the data, training a machine learning model, evaluating models and generating synthetic data. Since all computations are performed on the secret shares, the servers cannot infer the input values nor intermediate results, thus ensuring that MPC guarantees \textit{input privacy}. We refer to Appendix \ref{app:mpc} for details on threat models and on how to perform addition and multiplication with a replicated secret sharing scheme, as well as an overview of MPC primitives that we use in this paper.


\vspace{0.3em}
\noindent
\textbf{Differential Privacy (DP).} Consider two neighboring datasets $D$ and $D'$ that differ by a single record.\footnote{$D$ can be obtained by ether adding or removing a record in $D'$.} A randomized algorithm $F$ is said to be $(\epsilon,\delta)$-DP if, for any pair of neighboring datasets $D$ and $D'$ and for all subsets $O$ in the range of the output of $F$, it holds that $\mbox{Pr}(F(D) \in O) \leq e^{\epsilon} \cdot \mbox{Pr}(F(D') \in O) + \delta$ - where $\epsilon$ and $\delta$ are the privacy parameters that represent the privacy budget $\epsilon$ (measure of privacy loss), and the probability $\delta$ of the privacy being compromised, respectively \cite{dwork2014algorithmic}. The smaller the values of these parameters, the stronger the privacy guarantees. A DP algorithm $F$ is usually created out of an algorithm $F^*$ by adding noise proportional to the sensitivity of $F^*$, where sensitivity is maximum change in $F^*$'s output when comparing outputs on $D$ and $D'$. The \textit{Gaussian Mechanism} achieves this 
by adding noise drawn from a Gaussian (normal) distribution with $0$ mean and 
with a standard deviation scaled to the $l_2$-sensitivity of $F^*$.

\vspace{0.3em}
\noindent
\textbf{Synthetic Data Generation (SDG).} Consider a database $D$ with $d$ features (attributes) denoted by $x = \{x_1,x_2,\ldots,x_d\}$. The domain for  feature $x_i$ is a finite, discrete set represented by $\Omega_i$, i.e.,~$|\Omega_i| = \omega_i$. Let $\mathcal{Q}$ represent a collection of measurement sets where each set $q$ in $Q$ is a set of features to measure, i.e., $q \subseteq x$. 
A marginal on $q$ is denoted by $\mu_q(D)$ and simply refers to the number of occurrences in $D$ for each $t \in \Omega_q$ where $\Omega_q = \prod_{x_i \in q}\Omega_i$. Example: all 2-way marginals will consist of all $q$ where $|q|=2$, i.e., all 2-combinations of features such as $\{x_i,x_j\}$ where $i,j \in \{1,\ldots,d\}$ and $i<j$. Informally, one can for instance think of the 2-way marginal on $\{\textit{Gender},\textit{Age}\}$ as a histogram over all possible combinations of gender and age values in $D$. ``Select-measure-generate'' SDG algorithms aim at creating synthetic tabular data with marginals that are as close as possible to the original data \cite{hardt2010simple,zhang2017privbayes,mckenna2022aim,mckenna2021winning}.
In this paper, we use 
Private-PGM \cite{mckenna2019graphical} as a prototypical algorithm from this family of SDGs. Private-PGM constructs undirected graphical models from DP noisy measurements over low-dimensional marginals, which facilitates the generation of new synthetic samples via sampling from the learned graphical model. It works on records with discrete attributes. 
Simply put, for each $q$, Private-PGM firstly computes DP marginals $\mu_q(D) + \mathcal{N}(0, \sigma^2_q I)$, where $\mathcal{N(.)}$ is Gaussian noise with scale $\sigma^2_q$ determined based on $(\epsilon,\delta)$. 
This is the \textit{measurement step}. Then it estimates the joint marginal distribution that best explains all the noisy measurement. In parallel, it estimates the parameters of the graphical model using graph inference and learning algorithms such as belief propagation on a junction tree. This is the  \textit{generate step}. We refer to McKenna et al.~\cite{mckenna2019graphical} for more details.

%
%

\section{Secure Generation and Publishing of Synthetic Data}\label{sec:framework}
\textbf{Framework Setup.} We consider a scenario with $n$ data custodians $\mathbf{C} = \{C_1, C_2,\ldots, C_n\}$, who each hold a private dataset $D_i$. Their goal is to collaboratively generate synthetic data $\hat{D}$ of a desired quality using a process $F$ over the combined data $D = \bigcup_{i=1}^{n} D_i$, while preserving the privacy of the individual datasets, i.e., both input and output privacy. To achieve this, the data custodians utilize our proposed framework, which employs a set of $m$ non-colluding and independent MPC servers $\mathbf{S} = \{S_1, S_2,\ldots, S_m\}$ that perform computations over secret-shared data. These servers are equipped with MPC protocols to compute $F$, which comprises a privacy-preserving pipeline for hyperparameter tuning of the SDG model including MPC protocols for privacy-preserving preprocessing ($\pPRE$), privacy-preserving training of an SDG model ($\pSDG$) and privacy-preserving evaluation of generated synthetic data ($\pEVAL$).

\noindent
\textbf{Overview of the Framework.}
Algorithm \ref{alg:overview} provides an overview of our proposed framework. It iteratively searches for the hyperparameters of the SDG that yield the desired quality of synthetic data, employing K-fold cross-validation in each iteration. It implements the MPC protocols for major components of the SDG pipeline, taking privacy parameters $(\epsilon_p, \delta_p)$ and $(\epsilon_s, \delta_s)$ as input for preprocessing  the secret-shared data and training the SDG model  on secret-shared data, respectively. Only the components involved in releasing synthetic data -- preprocessing and training the SDG model (Lines 27, 28) -- are made DP, while the results of other components, such as evaluation, are never published and thus do not apply DP. Therefore, during K-fold cross-validation, we tune the parameters based on DP preprocessing and DP training (Lines 13,14), even though the output of this parameter tuning or K-fold cross-validation are not published.

\begin{algorithm}[ht!]
\caption{Privacy-preserving publishing of synthetic datasets}
\label{alg:overview}
\begin{algorithmic}[1]
\Require Set of MPC servers $\mathbf{S}$, Data custodians $\mathbf{C}$, number of folds K, 
privacy budget ($\epsilon_s, \delta_s$) for published synthetic data $\hat{D}$ and ($\epsilon_p, \delta_p$) for pre-processing, Maximum number of loops $L$, Set of hyperparameters to search for {$\mathbf{H}$}
\State Data custodians, $\mathbf{C}$,  secret share their respective datasets with MPC servers, $\mathbf{S}$,  as $[\![D_i]\!]$ and thresholds of data quality $[\![T_i]\!]$
\State $\mathbf{S}$ run setup 
\State $[\![D]\!] \leftarrow \pCOMB([\![D_i]\!] \,|\, \forall i \in |\mathbf{S}|)$ 
\State publish $\leftarrow$ false
\State max\_loops $\leftarrow 0$
\Repeat
\State max\_loops $\leftarrow$ max\_loops + 1
\State Get random indices for K-fold
\State Get hyperparameters $\mathbf{H_k}$ for training SDG
\State Initialize $\bold{R}$ of length K
\For{every fold k} 
\State $ [\![D_{train}]\!], [\![D_{test}]\!] \leftarrow $ getData($[\![D]\!]$)
\State $[\![D_{train}]\!]  \leftarrow \pPRE([\![D_{train}]\!],(\epsilon_p,\delta_p))$
\State $[\![\hat{D}_{train}]\!] \leftarrow \pSDG([\![D_{train}]\!], (\epsilon_s, \delta_s), \mathbf{H_k})$
\State $[\![r]\!] \leftarrow \pEVAL([\![\hat{D}_{train}]\!], [\![D_{test}]\!], [\![D_{train}]\!])$
\State $[\![R[k]]\!] \leftarrow [\![r]\!]$
\EndFor
\State $[\![M]\!] \leftarrow \pAVG([\![R[k]]\!] | \forall k)$
\State votes = $| \mathbf{C} |$
\For{every data custodian $c$ get threshold $[\![T_c]\!]$}
\State votes $\leftarrow$ votes $ - ([\![M]\! < [\![T_c]\!]))$
\EndFor
\State publish $\leftarrow \pEQ$ (votes, $| \mathbf{C} |$) 
\State  $\mathbf{H_s} \leftarrow \mathbf{H_k}$
\Until{publish or (max\_loops $\geq L$) or no $\mathbf{H_k}$}
\If{publish}
\State $[\![{D}]\!] \leftarrow \pPRE([\![D]\!], (\epsilon_p,\delta_p))$
\State $[\![\hat{D}]\!] \leftarrow \pSDG([\![D]\!], 
(\epsilon_s,\delta_s), \mathbf{H_s})$
\State $[\![\hat{D}]\!] \leftarrow \pREVPRE([\![\hat{D}]\!])$
\State Reveal $\hat{D}$ to $\mathbf{C}$
\EndIf
\end{algorithmic}
\end{algorithm}

The data custodians $\mathbf{C}$ secret share their private data $[\![D_i]\!]$ and the desired quality of synthetic data as thresholds $[\![T_i]\!]$ to MPC servers $\mathbf{S}$ on Line 1, while servers set up based on the secret-shared inputs received from the data custodians and public inputs such as the dimensions of the data and the set of parameters to tune for. $\mathbf{S}$ then run $\pCOMB$ on Line 3 to concat the secret shares of the data from all
$\mathbf{C}$ and hold the secret shares of $[\![D]\!]$ at the end of this protocol, hence, mimicking the centralized setup, as if all the data was in one location. 

Parameter tuning is performed iteratively from Lines 6 -- 25 until one of the following conditions is met: the hyperparameter set is exhausted, the maximum number of iterations is reached, or the desired quality of synthetic data is achieved.\footnote{This can be easily modified to exhaustively search for all parameters, and then choose the best hyperparameter for which we obtained the best evaluation metrics. This would require large number of secure comparisons - comparing with thresholds and comparing with all the sets of averaged evaluation metrics.} Each iteration begins with K-fold (Lines 8 -- 17), creating train and test splits based on indices only with getData() on Line 12. $\mathbf{S}$ then run MPC protocol $\pPRE$ to preprocess the secret shares of training data on Line 13 followed by an MPC protocol for $\pSDG$ to train an SDG model on secret shares of preprocessed training data with the set of hyperparameter $\mathbf{H_k}$. On line 15, $\mathbf{S}$ run $\pEVAL$ to compute secret shares of the pre-defined evaluation metrics $[\![r]\!]$ based on secret shares of both the real $[\![D_{train}]\!], [\!D_{test}]\!]$ and generated synthetic data $[\![\hat{D}_{train}]\!]$. The value of $[\![r]\!]$ is never published and hence need not be DP. Note that all of these steps perform computation on combined real data and hence are done in MPC. 
On line 18, $\mathbf{S}$ compute secret shares of K-fold evaluation $[\![M]\!]$ using $\pAVG$. 

Lines 19 -- 23 determine if $[\![M]\!]$ meets the thresholds of all the data custodians. To do so, assume that $[\![M]\!]$ meets the threshold for all $\mathbf{C}$ on Line 19, i.e., votes $= |\mathbf{C}| $. Line 21 decrements the votes if it does not meet the threshold of a data custodian, finally checking if the number of votes still equals $|\mathbf{C}|$ on Line 23. If yes, it is set to publish on Line 23, selecting the hyperparameters for this iteration as the final set of hyperparamters $\mathbf{H_s}$.  The voting is done in a privacy-preserving way -- (a) it does not require release of any evaluation metrics; and (b) the thresholds of each data custodian and the number of votes are kept confidential. 
Finally, on Line 27 -- 29, the SDG pipeline is run on the combined data $[\![D]\!]$, including reverse preprocessing on Line 29 to match the format of the real data that was secret shared initially. Line 30 publishes generated synthetic data of the same length as the combined data, which can be changed in the framework (e.g.~on Line 28). Note that the framework is modular in nature and can be switched to suit any specific preprocessing or SDG algorithm or evaluation metrics\footnote{The code for our framework is available at \url{https://github.com/sikhapentyala/MPC_SDG_param_tune}. Note that it is work in progress.}. Once we have the MPC protocols in-place, one can easily modify to change the method of parameter search as long as it does not require access to private data. 

%
%

\section{Use Case: Privacy-Preserving Publishing of Synthetic Genomic Data for Leukemia} \label{sec:alldata}
Sharing genomic data is essential for advancing biomedical research with AI; however, such data is highly privacy-sensitive and hard to access \cite{bonomi2020privacy}. To address this data bottleneck, researchers are exploring the potential of sharing synthetic datasets that protect patient privacy \cite{oestreich2021privacy}. More often than not, however, genomic data is controlled by multiple data custodians, for who sharing their data such as in the centralized paradigm with a central aggregator raises privacy concerns. 
Our proposed framework is designed for federated SDG in scenarios like these. Below, we demonstrate our \textit{ongoing research} on the use of our framework to publish synthetic genomic data for leukemia.

\textbf{Description of Data.} We consider bulk RNA-seq data compiled as a matrix, where each row represents the specimen of a patient and each column represents a gene expression level \cite{chen2024towards}.\footnote{This dataset is openly available on \url{https://github.com/MarieOestreich/PRO-GENE-GEN} and is used purely to demonstrate our solution.} It includes samples from 5 disease categories, 4 of which are types of leukemia {AML, ALL, CML, CLL}, while the fifth is classified as ``Other.'' The dataset has 1,181 samples and 12k genes, which is reduced to 958 genes based on the L1000 list.

\textbf{Generating Synthetic Genomic Data with Proposed Framework.} Generating synthetic data for genomic applications is particularly challenging. Below we use a select-measure-generate SDG method for tabular data, namely Private-PGM \cite{mckenna2019graphical}.\footnote{Our initial experiments with other SDG methods of the same ``select-measure-generate paradigm'' such as AIM \cite{mckenna2022aim} indicated poor performance. We are yet to explore other categories of SDGs and additional SDG algorithms. Nonetheless, the limited suitability of existing SDG algorithms for genomic data highlights the need for research into novel techniques of genomic SDG.}  This method takes categorical data as input and benefits from a small domain size for each feature. Our sample record is denoted by $g = \{g_1,g_2,\ldots,g_d,y\}$ where $g_i$ is the gene expression level for the $i$th gene and $y$ is the label that denotes the kind of leukemia. We compute all 1-way marginals as well as the 2-way marginals associated with the label attribute (i.e.  a total of 959 measurements with $|q|=1$ and 958 measurements with $|q|=2$, resulting in 1,917 noisy marginals). The privacy budget is allocated uniformly across each measurement i.e.~$\epsilon_q = \epsilon_s/1,917$ following sequential composition. To evaluate the quality of generated synthetic data, we train a logistic regression model to infer the leukemia type based on the gene expression values, and we compute the workload error for 1-way marginals. 

\subsection{Secure Preprocessing}
Chen et al.~\cite{chen2024towards} generate synthetic data based on the leukemia dataset, assuming that it is available in its entirety with one aggregator. They preprocess the data by binning each gene feature into intervals based on quantiles (0.25, 0.5, and 0.75). Each gene is then assigned one of the four values in $\{0,1,2,3\}$, thereby reducing the domain size to $\omega_i = 4$, $\forall i \in [1,d]$, while the label domain size $\omega_y$ remains $5$. Once the final synthetic data is generated, it is de-binned (reverse processed). De-binning replaces each bin with the computed mean of the original values within that bin.
We extend this approach to the cross-silo setting by proposing an MPC protocol $\pBIN$, which is a concrete instantion of $\pPRE$ from Algorithm \ref{alg:overview}. $\pBIN$ performs quantile binning across data silos while keeping the data encrypted \footnote{Please refer to Appendix \ref{app:bin} for note and our ongoing research on DP quantile binning}. De-binning is performed similarly with MPC protocol $\pINVBIN$, an instantiation of $\pREVPRE$ from Algorithm \ref{alg:overview}.

\begin{algorithm}[ht!]
    \caption{$\pBIN$: Protocol to bin the genomic data ($\pPRE$)}
\label{alg:pibin}
\begin{algorithmic}[1]
\Require Secret shares of one gene feature column of $[\![D^g]\!]$ for gene $g$, the number of samples $N$
\Ensure Secret shares of one gene feature binned column $[\![D^g]\!]$ for gene $g$, secret shares of mean $[\![m^g\!]$ for gene $g$ for inverse discretization done later
\State Set quantiles = [0.25,0.5,0.75] and initialize $Q$ of size 3
\State Create a copy of $[\![D^g]\!]$ as $[\![{D^g_{copy}}]\!]$; Initialize $[\![Q]\!]$ of length 3
\State $[\![\widetilde{D^g}]\!] \leftarrow \pSORT([\![D^g]\!])$ \Comment{MPC protocol for sorting based on radix sort; see MP-SPDZ \cite{cryptoeprint:2020:521}}
\For{each quantile $r$ in quantiles} \Comment{Compute three quantiles}
\State pos $\leftarrow (N-1) \cdot r$ 
; $f$ $\leftarrow$ pos $-$ $\lfloor$pos$\rfloor$
\State $[\![Q[$$r$$]]\!] \leftarrow [\![\widetilde{D^g}[\lfloor$pos$\rfloor]]\!]$ + $f \cdot$($[\![\widetilde{D^g}[\lfloor$pos$\rfloor + 1]]\!] $-$[\![\widetilde{D^g}[\lfloor$pos$\rfloor]]\!]$)
\EndFor
\For{$i\gets 1$ to $N$} \Comment{Binning data into 4 bins with values {0,1,2,3}}
\State $[\![c_0]\!] \leftarrow \pLT([\![D^g[i]]\!]$, $[\![Q[$0$]]\!])$;
$[\![c_1]\!] \leftarrow \pLT([\![D^g[i]]\!]$, $[\![Q[$1$]]\!])$;
$[\![c_2]\!] \leftarrow \pLT([\![D^g[i]]\!]$, $[\![Q[$2$]]\!])$
\State $[\![D^g[i]]\!] \leftarrow 3 - [\![c_0]\!] - [\![c_1]\!]- [\![c_2]\!]$
\EndFor
\State Initialize $[\![m^g]\!]$ and $[\![$ctr$^g]\!]$ each of length 4
\For{$i\gets 1$ to $N$} \Comment{Computing mean for inverse discretization}
\For{$b\gets 0$ to $3$}
\State $[\![c]\!] \leftarrow \pEQ([\![D^g[i]]\!], b)$
\State $[\![m^g[b]\!] \leftarrow [\![m^g[b]\!] + \pMUL([\![c]\!] , [\![D^g_{copy}[i]]\!]$) 
\State $[\![$ctr$^g[b]]\!] \leftarrow [\![$ctr$^g[b]]\!] + [\![c]\!]$
\EndFor
\EndFor
\For{$b\gets 0$ to $3$}
\State $[\![m^g[b]\!] \leftarrow \pDIV([\![m^g[b]\!],[$ctr$^g[b]]\!])$
\EndFor
\end{algorithmic}
\end{algorithm}



\noindent
\textbf{Description of $\pBIN$.} $\pBIN$ takes as input secret shares of a single gene feature $g$ as $[\![{D^g}]\!]$ and outputs the binned gene column along with the corresponding mean values $[\![m^g]\!]$ to be used during inverse preprocessing. Lines 3--7 in Algorithm \ref{alg:pibin} compute the secret shares of the quantiles (the cut points) in a straightforward manner by executing an MPC protocol $\pSORT$ to sort the secret shares of the $N$ samples of one gene (see Bogdanov et al.~\cite{bogdanov2013oblivious} for an overview of oblivious sorting algorithms). 
A secret-shared vector of quantiles  $[\![Q]\!]$ is computed following, for each $r$ in $[0.25, 0.50, 0.75]$: $pos = (N-1) \cdot r$ 
and 
$[\![Q[$$r$$]]\!] = [\![\widetilde{D^g}[\lfloor$pos$\rfloor]]\!]$ + $f \cdot$($[\![\widetilde{D^g}[\lfloor$pos$\rfloor + 1]]\!] $-$[\![\widetilde{D^g}[\lfloor$pos$\rfloor]]\!]$).
%
Once the secret shares of the cut-off points are computed, Lines 8–11 bin the data by comparing the encrypted value  $[\![{D^g[i]}]\!]$ with the encrypted threshold $[\![{Q[r]}]\!]$ for each quantile $r$, using an MPC protocol $\pLT$ for comparison that returns a secret sharing of 1 if the first argument is smaller than the second, and a secret sharing of 0 otherwise (see Appendix \ref{app:mpc}). In line 10, we compute the bin index by leveraging the small finite size of bins to minimize computationally heavy operations $g_i = 3 - (g_i<Q[0]) - (g_i<Q[1]) - (g_i<Q[2])$. 

Lines 13--22 compute the mean values for each bin index in a straightforward manner, using MPC protocols for equality testing $\pEQ$ and multiplication $\pMUL$ of secret-shared values (see Appendix \ref{app:mpc}).  Note that $[\![$ctr$^g[b]]\!]$ is used as a secret-shared accumulator variable for the number of values that fall into bin $b$. 

\noindent
\textbf{Description of $\pINVBIN$.} $\pINVBIN$ takes as input secret shares of binned $[\![{D}]\!]$ and secret shares of all mean values $[\![m^g]\!]$ and outputs the de-binned data. We replace the value $g_i$ with the mean following $\sum_{i=0}^3 m^g[b]\cdot(g_i ==b)$ on Lines 4--5. We note that 
line 4 can be further optimized by using the indicator polynomials described in Section \ref{sec:sdg}.




\begin{algorithm}[ht!]
    \caption{$\pINVBIN$: Protocol to inverse discretize genomic data ($\pREVPRE$)}
\label{alg:pirevbin}
\begin{algorithmic}[1]
\Require Secret shares of binned data of $[\![D]\!]$, the number of samples $N$, secret shares of mean $[\![m^g\!]$ for all genes
\Ensure Secret shares of de-binned data $[\![D]\!]$
\For{$i\gets 1$ to $N$} \Comment{Binned Data into 4 bins with values {0,1,2,3}}
\For{every gene $g$}
\State $[\![x]\!] \leftarrow [\![D[i][g]]\!]$
\State $[\![c_0]\!] \leftarrow \pEQ([\![x]\!], 0)$;
$[\![c_1]\!] \leftarrow \pEQ([\![x]\!], 1)$;
$[\![c_2]\!] \leftarrow \pEQ([\![x]\!], 2)$;
$[\![c_3]\!] \leftarrow \pEQ([\![x]\!], 3)$
\State $[\![D[i][g]]\!] \leftarrow \pMUL([\![c_0]\!] , [\![m^g\!][0]) + 
\pMUL([\![c_1]\!], [\![m^g\!][1]) + 
\pMUL([\![c_2]\!] , [\![m^g\!][2]) +  
\pMUL([\![c_3]\!] , [\![m^g\!][3])$
\EndFor
\EndFor
\end{algorithmic}
\end{algorithm}

\subsection{Secure SDG}\label{sec:sdg}
While MPC protocols for the measurement step in marginals-based SDG algorithms have been proposed in earlier works \cite{maddock2023flaim,pentyala2024caps,  ramesh2023decentralised}, here we propose an efficient protocol that leverages the predefined small domain size.
We denote 1-way marginals for gene $g_i$ as $\mu_{g_i}$ and the 1-way marginal for label $y$ as $\mu_{y}$. Given $\Omega_{g_i} = \{0,1,2,3\}$ and $\Omega_{y} = \{0,1,2,3,4\}$, to compute a marginal, we compute the number of occurrences for all the values of $\Omega_q$ in $D$ where $q \in \{g_1,\ldots ,g_d, y\}$. 
To do so, for example, we need to compare each value in the data column $D^{g_i}$ of gene $g_i$ to each value in its discretized domain $\Omega_{g_i}$. To avoid equality test operations,\footnote{Comparison operations such as ``equals to'' are computationally heavy operations in MPC.} we define indicator polynomials, denoted by $I$ as follows:

\begin{minipage}{0.6\textwidth}
$
\begin{array}{lcl}
I_0(x) = \frac{(1-x)(2-x)(3-x)}{6} & \phantom{xxx} &
I_1(x) = \frac{x(2-x)(3-x)}{2}\\
& & \\
I_2(x) = \frac{x(x-1)(3-x)}{2} & &
I_3(x) = \frac{x(x-1)(x-2)}{6} \\
\end{array}
$
\end{minipage}
\begin{minipage}{0.4\textwidth}
$$
\begin{array}{r|cccc}
x & 0 & 1 & 2 & 3 \\
\hline
I_0(x) & 1 & 0 & 0 & 0 \\
I_1(x) & 0 & 1 & 0 & 0 \\
I_2(x) & 0 & 0 & 1 & 0 \\
I_3(x) & 0 & 0 & 0 & 1
\end{array}
$$
\end{minipage}

\begin{algorithm}[ht!]
    \caption{$\pNM$: Protocol to compute noisy marginals for binned genomic data}
\label{alg:noisymarginal}
\begin{algorithmic}[1]
\Require Secret shares of binned data $[\![D]\!]$, scale $\sigma$, number of patients $N$, Domain of gene features $\Omega_g = {0,1,2,3}$ and classes $\Omega_y = {0,1,2,3,4}$
\Ensure Secret shares of noisy marginals $[\![\mu]\!]$ for 1-way and selected 2-way marginals
\State Initialize matrix $[\![\mu_g]\!]$ of size $[d,4]$, array $[\![\mu_y]\!]$ of size $[5]$ and matrix $[\![\mu_{g,y}]\!]$ of size $[d,20]$
\For{$i\gets 1$ to $N$}

\State $[\![x]\!] \leftarrow [\![D[i][y]]\!]$ \Comment{Compute 1-way marginal for the label}
\State Initialize array $[\![L]\!]$ of size 5
\State $[\![s_0]\!] \leftarrow [\![x]\!]$; $[\![s_1]\!] \leftarrow [\![x]\!] - 1$; $[\![s_2]\!] \leftarrow [\![x]\!] - 2$; $[\![s_3]\!] \leftarrow [\![x]\!] -3$; $[\![s_4]\!] \leftarrow [\![x]\!] -4$; 

\State $[\![L[0] ]\!] \leftarrow \pMUL([\![s_1]\!] , [\![s_2]\!] , [\![s_3]\!] , [\![s_4]\!] , (1/24))$
\State $[\![L[1] ]\!] \leftarrow \pMUL([\![s_0]\!] , [\![s_1]\!] , [\![s_3]\!] , [\![s_4]\!] , (1/6))$
\State $[\![L[2] ]\!] \leftarrow \pMUL([\![s_0]\!] , [\![s_1]\!] , [\![s_2]\!] , [\![s_4]\!] , (1/2))$
\State $[\![L[3] ]\!] \leftarrow \pMUL([\![s_0]\!] , [\![s_1]\!] , [\![s_2]\!] , [\![s_4]\!] , (1/6))$
\State $[\![L[4] ]\!] \leftarrow \pMUL([\![s_0]\!] , [\![s_1]\!] , [\![s_2]\!] , [\![s_3]\!] , (1/24))$

\For{$r \in \Omega_y$}
\State $[\![\mu_y][r]\!] \leftarrow [\![\mu_y][r]\!] + [\![L[r] ]\!]$
\EndFor

\For{every gene $g$} \Comment{Compute 1-way marginals for all genes}
\State $[\![x]\!] \leftarrow [\![D[i][g]]\!]$
\State Initialize array $[\![G]\!]$ of size 5
\State $[\![s_1]\!] \leftarrow 1- [\![x]\!]$; $[\![s_{11}]\!] \leftarrow [\![x]\!] - 1$;
$[\![s_2]\!] \leftarrow 2- [\![x]\!]$; $[\![s_{21}]\!] \leftarrow [\![x]\!] - 2$; $[\![s_{3}]\!] \leftarrow 3 - [\![x]\!]$;
$[\![s_{31}]\!] \leftarrow [\![x]\!] - 3$;

\State $[\![G[0] ]\!] \leftarrow \pMUL([\![s_1]\!] , [\![s_2]\!] , [\![s_3]\!], (1/6))$
\State $[\![G[1] ]\!] \leftarrow \pMUL([\![x]\!] , [\![s_2]\!] , [\![s_3]\!] , (1/2))$
\State $[\![G[2] ]\!] \leftarrow \pMUL([\![x]\!] , [\![s_{11}]\!] , [\![s_3]\!] , (1/2))$
\State $[\![G[3] ]\!] \leftarrow \pMUL([\![x]\!] , [\![s_{11}]\!] , [\![s_{22}]\!] , (1/6))$

\For{$r \in \Omega_g$}
\State $[\![\mu_g][r]\!] \leftarrow [\![\mu_g][r]\!] + [\![G[r] ]\!]$
\EndFor

\For{$r \in \Omega_g$} \Comment{Compute 2-way selected marginals}
\State idx $\leftarrow r*5$ 
\For{$f \in \Omega_y$}
\State $[\![\mu_{g,y}[g][idx+f] ]\!] \leftarrow [\![\mu_{g,y}[g][idx+f] ]\!] + \pMUL([\![G[r] ]\!],[\![L[f] ]\!]) $ 
\EndFor
\EndFor

\EndFor

\EndFor

\For{every value $[\![x]\!]$ in $[\![\mu_g]\!]$, $[\![\mu_l]\!]$, $[\![\mu_lg]\!]$} \Comment{Adding noise to guarantee DP} 
\State $[\![x]\!] \leftarrow [\![x]\!] + \sigma \cdot \pGUASS(0,1) $
\EndFor 

\end{algorithmic}
\end{algorithm}

Using the above equations, we compute $\mu_{g_i}^{b} = \sum_{j=1}^N I_{b}(D[j][g_i])$, for $b \in \{0,1,2,3\}$.



The technique with indicator polynomials also works for the label feature with discretized domain $\Omega_{y} = \{0,1,2,3,4\}$: 

\begin{minipage}{0.6\textwidth}
$
\begin{array}{lcl}
I'_0(y) = \frac{(y-1)(y-2)(y-3)(y-4)}{24} &  \phantom{xxx} &
I'_1(y) = \frac{y(y-2)(y-3)(y-4)}{6} \\
& & \\
I'_2(y) = \frac{y(y-1)(y-3)(y-4)}{2} & &
I'_3(y) = \frac{y(y-1)(y-2)(y-4)}{6} \\
& & \\
I'_4(y) = \frac{y(y-1)(y-2)(y-3)}{24} & & \\
\end{array}
$
\end{minipage}
\begin{minipage}{0.4\textwidth}
$$
\begin{array}{r|ccccc}
y & 0 & 1 & 2 & 3 & 4\\
\hline
I'_0(y) & 1 & 0 & 0 & 0 & 0 \\
I'_1(y) & 0 & 1 & 0 & 0 & 0\\
I'_2(y) & 0 & 0 & 1 & 0 & 0\\
I'_3(y) & 0 & 0 & 0 & 1 & 0 \\
I'_4(y) & 0 & 0 & 0 & 0 & 1
\end{array}
$$
\end{minipage}

Then, $\mu_{y}^{b} = \sum_{j=1}^N I'_{b}(D[j][y])$ for ${b} \in \{0,1,2,3,4\}$. 

We leverage the precomputed polynomials above to compute 2-way marginals $\mu_{g_i,y}$. We further optimize this by considering each 2-way marginal to be a flattened array of fixed size $20 = 4\times5$. 
Then 
$$\mu_{g_i,y}^{b_g,b_y} = \sum_{j=1}^N I_{b_g}^{g_i}(D[j][g_i]) \cdot I_{b_y}^l(D[j][y]) \mbox{\ for\ } b_g \in \{0,1,2,3\} \mbox{\ and\ } b_y \in \{0,1,2,3,4\}$$

Protocol \ref{alg:noisymarginal} computes the secret shared noisy marginals. Lines 2--32 for computing secret shares of 1-way and 2-way marginals are straightforward, as they directly follow the equations described above. Once the marginals are computed, Gaussian noise is added to each of them. We utilize existing MPC protocols to generate Gaussian noise on Lines 33--35 (see Appendix \ref{app:mpc} in \cite{pentyala2024caps}). An MPC protocol for the generate step is future work.

\subsection{Secure Evaluation of Synthetic Data}\label{sec:eval}
We evaluate the quality of synthetic data based on two metrics -- (a) the accuracy of a logistic regression model trained for multiclass classification to identify the disease type, and (b) workload error. As instantiations of $\pEVAL$ on line 15 in Algorithm \ref{alg:overview}, these evaluations are to be performed while keeping all data encrypted, hence we need MPC protocols to execute them. 

For (a) we use an MPC protocol for training and inference with logistic regression, building upon the MPC primitives available for Dense and Softmax layers in MP-SPDZ \cite{cryptoeprint:2020:521}.  
For (b), we designed an MPC protocol $\pWLE$ that computes secret shares of the workload error following the equation 
$$e = \frac{1}{|\mathcal{Q}|} \sum_{q \in \mathcal{Q}} \sum_{b \in \Omega_q} |\mu_q^b(D) - \mu_q^b(\hat{D})|$$
Protocol \ref{alg:wle}  computes $\mu$ following the logic in Protocol \ref{alg:noisymarginal} and the rest of the computations are straightforward.\footnote{Our current implementation does not yet optimize to leverage the precomputed marginals for computation of error; this relatively straightforward step is left as future work.} 

\begin{algorithm}[ht!]
    \caption{$\pWLE$: Protocol to compute normalized marginal error over 1-way marginals ($\pEVAL$)}
\label{alg:wle}
\begin{algorithmic}[1]
\Require Secret shares of binned data $[\![D]\!]$, Secret shares of binned synthetic data $[\![\hat{D}]\!]$,  number of patients $N$, Domain of gene features $\Omega_g = {0,1,2,3}$ and classes $\Omega_y = {0,1,2,3,4}$
\Ensure Secret shares of workload error $[\![e]\!]$ 
\State Initialize $[\![e]\!]$
\State $[\![\mu_g]\!]$ , $[\![\mu_y]\!]$ , $[\![\mu_{g,y}]\!] \leftarrow$ Run Lines 1--32 from Protocol \ref{alg:noisymarginal} on $[\![D]\!]$

\For{every value $[\![x]\!]$ in $[\![rc_g]\!]$, $[\![rc_l]\!]$, $[\![rc_lg]\!]$} \Comment{Adding noise to guarantee DP} 
\State $[\![x]\!] \leftarrow \pMUL([\![x]\!], (1/N)) $
\EndFor 

\State $[\![\hat{\mu}_g]\!]$ , $[\![\hat{\mu}_y]\!]$ , $[\![\hat{\mu}_{g,y}]\!] \leftarrow$ Run Lines 1--32 from Protocol \ref{alg:noisymarginal} on $[\![\hat{D}]\!]$

\For{every value $[\![x]\!]$ in $[\![\hat{\mu}_g]\!]$, $[\![\hat{\mu}_l]\!]$, $[\![\hat{\mu}_lg]\!]$}
\State $[\![x]\!] \leftarrow \pMUL([\![x]\!], (1/N)) $
\EndFor

\For{(every value $[\![\hat{x}]\!]$ in $[\![\hat{\mu}_g]\!]$, $[\![\hat{\mu}_l]\!]$, $[\![\hat{\mu}_lg]\!]$) and ($[\![x]\!]$ in $[\![{\mu}_g]\!]$, $[\![{\mu}_l]\!]$, $[\![{\mu}_lg]\!]$)
 }
\State $[\![e]\!] \leftarrow [\![e]\!] + \pABS([\![x]\!] - [\![\hat{x}]\!]) $
\EndFor

\end{algorithmic}
\end{algorithm}

We note that Line 21 in Algorithm \ref{alg:overview} decrements $[\![c]\!]$ for a data custodian  only if both the metrics meet the given thresholds. The data custodians will secret share an array of thresholds for each metric to this end.

\subsection{Empirical Evaluation}
We implemented our framework in MP-SPDZ \cite{keller2020mp} and leverage the MPC primitives available therein. We consider the number of iterations in the generate step of the Private-PGM as the hyperparameter to be tuned. For simplicity we consider $\mathbf{H}=\{10,15,25,30\}$.

\textbf{Utility evaluation.} We evaluate the framework in Algorithm \ref{alg:overview} on the genomic data for leukemia\footnote{For the results in this section, we include the ``generate'' step in the clear.}. 
Table \ref{tab:quality} reports the  quality of synthetic data that is published on line 30 in Algorithm \ref{alg:overview} when compared with the real data. While throughout the execution of Algorithm \ref{alg:overview}, we use MPC protocols to compute the workload error and to train and test logistic regression models for $\pEVAL$ on line 15, in Table \ref{tab:quality} we report the workload error (WLE) as defined in Section \ref{sec:eval} and the accuracy of a logistic regression and a decision tree model computed in the clear, i.e.~without MPC. This allows to assess the fidelity and the utility of the generated synthetic data without the influence of MPC.

For this experiment, we split the leukemia dataset into train and test datasets. We spilt the train dataset further into 2 equal parts, simulating a scenario with 2 data custodians $|\mathbf{C}| = 2$, where each data custodian holds a part of the train dataset. Algorithm \ref{alg:overview} takes the split train datasets as input, while the test dataset is kept aside for evaluation. 
We run our framework with $|S|=3$. Once the synthetic data is published at the end of Algorithm \ref{alg:overview}, we train a logistic regression model and a decision tree on the real train data and the generated synthetic data independently. We then evaluate each model on the test data, and report the absolute difference between accuracy and F1 score in Table \ref{tab:quality}.
To demonstrate the effectiveness of preprocessing across data silos, we conduct the experiment where data custodians perform binning (preprocessing) independently on their respective portions of the split data and then secret share, rather than preprocessing the combined dataset (row 1 of Table \ref{tab:quality}). Our experiments show that this dataset benefits from combined preprocessing for ML tasks, specifically the ML model and metric for which it was tuned (LR). The evaluation suggests that we need better SDG algorithms for genomic data. We further note that we considered a relatively simple format of genomic data.





\begin{table}[ht!]
 \caption{\textbf{Quality evaluation of the generated synthetic genomic data}. We run the framework with $|S|=3$. We set $k=5$ and tune the number of iterations of the Private-PGM model. Once the final synthetic data is published, we compute WLE between synthetic and combined real data and report difference of machine learning utility scores (accuracy and F1 scores with logistic regression $\Delta$LR and decision tree $\Delta$DT) between synthetic and combined real data  (note: this evaluation is not part of the framework). We report averages taken over 3 runs.}
   \centering
  \begin{tabular}{l r r r r r}
    \toprule
     & WLE & |$\Delta$LR Accuracy| & |$\Delta$LR F1| & |$\Delta$DT Accuracy| & |$\Delta$DT F1|\\
    \midrule
     Local preprocessing & \textbf{0.015} & 0.026 & 0.141 & 0.346 & 0.170\\
     Ours & 0.019 & \textbf{0.003} & \textbf{0.135} & \textbf{0.079} & \textbf{0.125}\\
    \bottomrule
  \end{tabular}
  \label{tab:quality}
\end{table}

\textbf{Performance evaluation.}  We run our framework for different threat models \cite{araki2016high, dalskov2021fantastic,cryptoeprint:2018:482} and  report the benchmarking runtimes for running the MPC protocols for leukemia data in Table \ref{tab:time}. We run for the similar setup as mentioned above, and therefore the input data size $N = 945$. For 3PC passive, the parts of the framework with MPC protocol run for a total of $\sim2.5$ hours -- one run of $\pBIN$ takes 82.93 secs., $\pINVBIN$ takes 11.83 secs., $\pNM$ takes 249.50 secs. ,$\pLR$ with 150 epochs takes 15.42 secs. and $\pWLE$ takes 158.66 secs. Table \ref{tab:time} reports performance evalaution for different threat models. The observed runtimes are in-line with the literature and demonstrate that an MPC based framework is feasible for end-to-end synthetic data generation where privacy is of utmost importance compared to the computational time required to generate valuable synthetic data. Future work will focus on optimizing these protocols and proposing new optimized MPC protocols for the proposed framework.  

\begin{table}[ht!]
 \caption{\textbf{Performance evaluation}. We run experiments with $|S|=2,3,4$ for different threat models with $\epsilon=5$ and $N = 945$. We report runtimes in seconds and total communication cost (Comm.) in GB  for one run of the major MPC protocols of our framework. 
 The performance is for the online phase of the MPC protocols. The runtimes are averaged over 5 runs. The experiments were run on F32 Azure VMs.}
  \centering
  \scalebox{0.92}{
    \small{
  \begin{tabular}{ll rr rr rr rr}
    \toprule
     & Protocol & \multicolumn{2}{c}{2PC Passive} & \multicolumn{2}{c}{3PC passive} & \multicolumn{2}{c}{3PC active} & \multicolumn{2}{c}{4PC active}\\
     &  & Time(s) & Comm(GB) & Time(s) & Comm(GB) & Time(s) & Comm(GB) & Time(s) & Comm(GB)\\
     \midrule
     $\pPRE$  & $\pBIN$   & 18259.66 & 2621 & 82.93 & 6.94 & 492.50 & 30.17 & 1174.44 & 88.90\\   
              & $\pINVBIN$  & 886.18 & 107 & 11.83 & 1.65 & 117.47 & 9.166 & 35.06 & 4.08\\    
   $\pSDG$    &  $\pNM$ & 3034.80 & 92.61 & 249.50 & 45.50 & 1664.42 & 348.74 & 357.23 & 119.71\\
   $\pEVAL$   &  $\pLR$ (150 ep) & 243.96 & 19.79 & 15.42 & 3.65 & 159.78 & 28.45 & 33.17 & 9.66\\
   &            $\pLR$ (300 ep) & 468.71 & 11.65 & 28.74 & 7.27 & 405.83 & 75.19 & 61.79 & 19.24\\
     &  $\pWLE$ & 2787.17 & 71.68 & 158.66  & 46.95 & 1799.66 & 499.05 & 329.86 & 123.86 \\
    \bottomrule
  \end{tabular}
  }
  }
  \label{tab:time}
\end{table}

In our experiments, we observed that if the data is binned without computing means required for debinning, then only the test data in each fold needs to be binned (example for $|d|=5$, the runtime of $\pBIN$ improves from 0.40 to 0.34). This improves the performance of individual protocols and may improve the overall framework's performance, particularly for high-dimensional data. We proposed protocols that can be combined in various ways to optimize performance or adapt to specific requirements. For $\pLR$, we report performance with 150 and 300 epochs illustrating that the MPC protocol scales linearly. 



\section{Conclusion}
We proposed an end-to-end framework for a synthetic data generation pipeline that produces synthetic data based on real data across multiple silos while ensuring both input and output privacy. Input privacy is achieved through MPC, and output privacy is ensured using DP. The framework incorporates MPC protocols for data preprocessing, hyperparameter tuning of the SDG model, and evaluation of the generated synthetic data, all while preserving privacy and within MPC. 

We evaluate our framework by generating synthetic genomic data for leukemia and propose the necessary MPC protocols to instantiate our framework to this end. Our evaluation demonstrates the feasibility of the proposed approach. However, generating high-quality synthetic genomic data remains an open challenge, even in the clear, and further research is needed to develop optimized MPC protocols.



\section*{Acknowldgements}
This material is based upon work supported by the National Science Foundation under Grant No.~2451163,  and by  NSF NAIRR 240091 (TACC Frontera). This research was, in part, funded by the National Institutes of Health (NIH) Agreement No. 1OT2OD032581. The views and conclusions contained in this document are those of the authors and should not be interpreted as representing the official policies, either expressed or implied, of the NIH.

\bibliographystyle{unsrt}
\bibliography{references}

\newpage
\appendix

\section{Secure Multi-Party Computation}\label{app:mpc}
We now give a brief introduction to MPC where we closely follow Section 2 in \cite{pentyala2024caps}.
\textit{Data Representation in MPC.} MPC works for inputs defined over a finite field or ring. Since inputs for synthetic data set generation algorithms are finite precision real numbers, we convert all of our inputs to fixed precision \cite{FC:CatSax10} and map these to integers modulo $q$, i.e.,~to the ring $\mathbb{Z}_q =\{0,1,\ldots,q-1\}$, with $q$ a power of 2. In fixed-point representations with $a$ fractional bits, each multiplication generates $a$ additional fractional bits. To securely remove these extra bits, we use the deterministic truncation protocol by \cite{dalskov2019secure}. 

\textit{Replicated sharing-based 3PC.} To give the reader an understanding of how MPC protocols work, we give a brief description of a specific MPC protocol based on replicated secret sharing with three computing parties \cite{araki2016high}. A private value $x$ in $\mathbb{Z}_q$ is secret shared among servers (parties) $S_1, S_2,$ and $S_3$ by selecting uniformly random shares $x_1, x_2, x_3 \in \mathbb{Z}_q$ such that 
$x_1 + x_2 +x_3 =  x \mod{q}$,
and giving $(x_1,x_2)$ to $S_1$, $(x_2,x_3)$ to $S_2$, and $(x_3,x_1)$ to $S_3$. We denote a secret sharing of $x$ by $[\![x]\!]$. 

From now on, servers will compute on the secret shares of $x$ rather than on $x$ itself. In order to proceed with the computation of a function, we need a representation of such function as a circuit consisting of addition and multiplication gates. The servers will compute the function gate by gate, by using specific protocols for computing addition of a publicly known constant to a secret shared value, addition of two secret shared values, multiplication of a secret shared value times a public constant, and multiplication of two secret shared values. After all the gates representing the function have been evaluated, the servers will hold secret shares of the desired final result of the computation. 

The three servers are capable of performing operations such as the addition of a constant, summing of secret shared values, and multiplication by a publicly known constant by doing local computations on their respective shares. 

To multiply secret shared values $[\![x]\!]$ and $[\![y]\!]$, we can use the fact that $x \cdot y=(x_1 + x_2 +x_3)(y_1 + y_2 +y_3) \mod{q}$. This means that $S_1$ computes $z_1=x_1 \cdot y_1+x_1 \cdot y_2+x_2 \cdot y_1 \mod{q}$, $S_2$ computes $z_2=x_2 \cdot y_2+x_2 \cdot y_3+x_3 \cdot y_2 \mod{q}$, and $S_3$ computes $z_3=x_3 \cdot y_3+x_3 \cdot y_1+x_1 \cdot y_3 \mod{q}$. The next step is for the servers to obtain an additive secret sharing of $0$ by choosing random values $u_1,u_2,u_3$ such that $u_1 + u_2 +u_3 = 0 \mod{q}$. This can be done using pseudorandom functions. Each server $S_i$ then computes $v_i=z_i+u_i \mod{q}$. Finally, $S_1$ sends $v_1$ to $S_3$, $S_2$ sends $v_2$ to $S_1$, and $S_3$ sends $v_3$ to $S_2$. This allows the servers $S_1, S_2$, and $S_3$ to obtain the replicated secret shares $(v_1,v_2)$, $(v_2,v_3)$, and $(v_3,v_1)$, respectively, of the value $v=x \cdot y$.


This protocol can be proven secure against honest-but-curious adversaries. In such case, corrupted players follow the protocol instructions but try to obtain as much knowledge as possible about the secret inputs from the protocol messages, and locally stored information. We can adapt this protocol to be secure even in the case of malicious adversaries. Those can arbitrarily deviate from the protocol in order to break its privacy. For the malicious case, we use the MPC scheme  proposed by \cite{dalskov2021fantastic} as implemented in MP-SPDZ \cite{cryptoeprint:2020:521}.

\subsection{Threat Models} \label{app:threat models}
MPC protocols are designed to protect against adversaries who corrupt one or more parties to either learn private information or manipulate computation results. We consider a static adversary model. In the \textit{semi-honest} (or \textit{passive}) model, even corrupted parties (MPC servers) follow the protocol but can attempt to extract private information from their internal states and received messages. MPC protocols that are secure against semi-honest adversaries prevent such information leakage. In the \textit{malicious} (or \textit{active}) model, corrupted parties can deviate arbitrarily from the protocol. MPC schemes that are secure against malicious adversaries ensure that such attacks are not successful, but this incurs higher computational costs than security against passive adversaries.

\subsection{MPC Primitives} 
The MPC schemes listed in Section \ref{sec:prelims} provide a mechanism for the servers to perform cryptographic primitives through the use of secret shares, namely addition of a constant, multiplication by a constant, and addition of secret shared values, and multiplication of secret shared values (denoted as $\pMUL$). Building on these cryptographic primitives, MPC protocols for other operations have been developed in the literature. 
We use \cite{cryptoeprint:2020:521}:
\begin{itemize}[leftmargin=*,noitemsep,topsep=0pt]

    \item Secure random number generation from uniform distribution $\pRDM$ : In $\pRDM$, each party generates $l$ random bits, where $l$ is the fractional precision of the power 2 ring representation of real numbers, and then the parties define the bitwise XOR of these $l$ bits as the binary representation of the random number jointly generated.

    \item Secure multiplication $\pMUL$ : At the start of this protocol, the parties have secret sharings $[\![a]\!]$ and $[\![b]\!]$; at the end, then they have a secret share of $c=a .b$. 
 
    \item Secure equality test $\pEQ$ : At the start of this protocol, the parties have secret sharings $[\![a]\!]$; at the end if $a = 0$, then they have a secret share of $1$, else a secret sharing of $0$.

    \item Secure less than test $\pLT$ : At the start of this protocol, the parties have secret sharings $[\![a]\!]$ and $[\![b]\!]$ of integers $a$ and $a$; at the end of the protocol they have a secret sharing of $1$ if $a < b$, and a secret sharing of $0$ otherwise.

     \item Secure less than test $\pABS$ : At the start of this protocol, the parties have secret sharings $[\![a]\!]$; at the end of the protocol they have a secret sharing of of its absolute value.
    
    \item Other generic subprotocols : $\pMUL$ also computes the product of secret shares of a given set of values. $\pSOFTMAX$ computes the softmax for a given vector. (See \cite{keller2022secure})
\end{itemize}
MPC protocols can be mathematically proven to guarantee privacy and correctness. We follow the universal composition theorem that allows modular design where the protocols remain secure even if composed with other or the same MPC protocols \cite{canetti2000security}.

\subsection{Protocol for Gaussian Noise}
Protocol \ref{alg:gs} samples the secret shares of the Gaussian distribution using $\mathcal{N}(0,1)$ Irwin-Hall approximation \cite{pentyala2024caps}. There are references to MPC protocols using Box-Mullers method in   \cite{pentyala2024caps} and other related work.
\begin{algorithm}[ht!]
    \caption{$\pGUASS$: Protocol to sample from Gaussian distribution $\mathcal{N}(0,1)$ }
\label{alg:gs}
\begin{algorithmic}[1]
\Ensure Secret shares of drawn sample $[\![\gamma]\!]$ 
\State Initialize vector $[\![\gamma]\!]$ of length $\max(\omega_r)$ with $0$s
   \For{i = 0 to $\omega_{q_s}$}
   \State $[\![sum]\!] \leftarrow 0$
   \For{j = 0 to 12}
   \State $[\![sum]\!] \leftarrow [\![sum]\!] + \pRDM(0,1)$
   \EndFor
   $[\![\gamma[i]]\!] \leftarrow [\![sum]\!] - 6$
   \EndFor
\end{algorithmic}
\end{algorithm}

\subsection{Note on DP implementation in MPC}
Keeping in mind the dangers of implementing DP with floating point arithmetic \cite{mironov2012significance}, we stick with the best practice of using fixed-point and integer arithmetic as recommended by, for example, OpenDP \footnote{\url{https://opendp.org/}}. We implement all our DP mechanism using their discrete representations and use 32 bit precision to ensure correctness.

We also remark that finite precision issues can also impact the exponential mechanism. If the utility of one of the classes in the exponential mechanism collapses to zero after the mapping into finite precision, pure DP becomes impossible to achieve. We can deal with such situation by using approximate DP for an appropriate value of $\delta$. Such situation did not happen with the data sets used in our experiments.

\subsection{DP Quantile Binning}\label{app:bin}
Ideally, we need to perform DP quantile binning. We closely follow \cite{chen2024towards} and consider this as our next steps.
Although various DP quantile estimation algorithms exist in the literature \cite{gillenwater2021differentially,kaplan2022differentially,durfee2023unbounded}, our initial results indicate that DP quantiles affects the utility for lower values of $\epsilon$ (i.e., higher privacy levels).
We are still exploring optimal DP quantile methods and alternative preprocessing approaches that are effective for genomic data \footnote{We are also investigating data structures, such as t-digest. \cite{dunning2021t}, which may require additional processing by data custodians. This is part of our ongoing research for effective preprocessing techniques.} Additionally, we looked into existing MPC protocols for quantile computation \cite{lan2023efficient,bohler2020secure,aseeri2021secqsa}. While these protocols are promising, they typically rely on approximate algorithms or require additional processing by data custodians, and do not ensure differential privacy. For our ongoing research, we designed a (naive) MPC protocol $\pBIN$ that computes exact quantiles, though it is not yet DP.

\end{document}